\documentclass{aa} 
\usepackage{amsmath}
\usepackage{graphicx}
\usepackage{color}
\usepackage{txfonts}
\usepackage[authoryear]{natbib}

\begin{document}
\title{Long-term change in the cyclotron line energy in Hercules X-1} 

\author{R.~Staubert\inst{1}, D.~Klochkov\inst{1}, J.~Wilms\inst{2},  K.~Postnov\inst{3}, 
N.I.~Shakura\inst{3}, R.E.~Rothschild\inst{4}, F.~F\"urst\inst{5}, F.A.~Harrison\inst{5}}

\offprints{staubert@astro.uni-tuebingen.de}

\institute{
	Institut f\"ur Astronomie und Astrophysik, Universit\"at T\"ubingen,
	Sand 1, 72076 T\"ubingen, Germany
\and
        Dr.\ Remeis Sternwarte, Astronomisches Institut der
	Universit\"at Erlangen-N\"urnberg, Sternwartstr. 7, 96049 Bamberg, Germany
\and
	Moscow M.V. Lomonosov State University, Sternberg Astronomical Institute, 119992, Moscow, Russia
\and
        Center for Astrophysics and Space Sciences, University of
        California at San Diego, La Jolla, CA 92093-0424, USA
\and
        Cahill Center for Astronomy and Astrophysics, California Institute of Technology, 
        Pasadena, CA 91125, USA 
}

\date{received: 2014 May 14, accepted: 2014 Oct 08}
\authorrunning{Staubert et al.}
\titlerunning{Long-term change in the cyclotron line energy in Her X-1}

\abstract
   {}
   {
We investigate the long-term evolution of the Cyclotron Resonance Scattering 
Feature (CRSF) in the spectrum of the binary X-ray pulsar \object{Her~X-1} 
and present evidence of a true long-term decrease in the centroid energy E$_{\rm cyc}$
of the cyclotron line in the pulse phase averaged spectra from 1996 to 2012.
   }
   {
Our results are based on repeated observations of Her~X-1 by
those X-ray observatories capable of measuring clearly beyond the 
cyclotron line energy of $\sim40$\,keV.
   }
   {
The historical evolution of the pulse phase averaged CRSF centroid energy 
E$_{\rm cyc}$ since its discovery in 1976 is characterized by an initial value 
around 35\,keV, an abrupt jump upwards to beyond $\sim40$\,keV
between 1990 and 1994, and an apparent decay
thereafter. Much of this decay, however, was found to be due to an
artifact, namely a correlation between E$_{\rm cyc}$ and the X-ray luminosity
$L_{x}$ discovered in 2007.
In observations after 2006, however, we now find a statistically significant true
decrease in the cyclotron line energy. At the same time, the dependence of E$_{\rm cyc}$ 
on X-ray luminosity is still valid with an increase of $\sim$ 5\% in energy for a factor of 
two increase in luminosity. 
A decrease in E$_{\rm cyc}$
by 4.2\,keV over the 16 years from 1996 to 2012 can either be modeled by a
linear decay, or by a slow decay until 2006 followed by a more abrupt decrease
thereafter. 
   }
   {
We speculate that the physical reason could be connected to 
a geometric displacement of the cyclotron resonant scattering region in the polar field or to a true 
physical change in the magnetic field configuration at the polar cap by the continued accretion. 
   }
  {}

   
   \maketitle
%

\section{Introduction}
\label{sec:introduction}

The accreting binary X-ray pulsar Her X-1 shows strong variability on several different timescales: 
the 1.24\,s spin period of the neutron star, the 1.7-day binary period, the 35-day period,
and the 1.65-day period of the pre-eclipse dips. The 35-day ON-OFF modulation 
can be understood as due to the precession of a warped accretion disk. Because of the 
high inclination ($i>80^\circ$) of the binary we see the disk nearly edge-on. The precessing 
warped disk therefore covers the central X-ray source during a substantial portion of the 
35-day period. Furthermore, a hot X-ray heated accretion disk corona reduces the X-ray 
signal (energy independently) by Compton scattering whenever it intercepts our line of sight 
to the neutron star. As a result the X-ray source is covered twice during a 35-day cycle.
Another 35\,d modulation is present in the systematic variation of the shape of the 1.24\,s
pulse profile. It has been suggested \citep{Truemper_etal86} that the reason for 
this is free precession of the neutron star, leading to a systematic change in our viewing angle to
the X-ray emitting regions. By comparing the variation in flux (the turn-ons) and the variations
in pulse shape, \citet{Staubert_etal09} had concluded, that if precession of the NS exists a strong interaction 
between the NS and the accretion disk is required, allowing a synchronization of the two clocks 
to nearly the same frequency through a closed loop physical feed-back (for which there is 
independent evidence).  
Further analysis of the variations in pulse profiles \citep{Staubert_etal13}, however, has shown 
that the histories of the turn-ons and of the variations in pulse shape are identical, with correlated 
variations even on short timescales ($\sim300$\,d). 

It is believed that the X-ray spectrum emerges from the hot regions
around the magnetic poles where the accreted material is channeled
by the $\sim10^{12}$\,G magnetic field down to the surface of the NS.
The height of the accretion mound is thought to be a few hundred meters. 
If the magnetic and spin axes of the neutron star are not aligned, the view of a terrestrial
observer is modulated at the rotation frequency of the star.

The X-ray spectrum of Her~X-1 is characterized by a power law continuum 
with exponential cut-off and an apparent line-like feature. The continuum is
believed to be due to thermal bremsstrahlung radiation from the $\sim$10$^{8}$\,K
hot plasma modified by Comptonization \citep{BeckerWolff_07,Becker_etal12}.
The line feature was discovered in 1976 in a balloon observation \citep{Truemper_etal78}.
This feature is now generally accepted as an absorption feature around
40\,{\rm keV} due to resonant scattering of photons off electrons on
quantized energy levels (Landau levels) in the Terragauss magnetic
field at the polar cap of the neutron star. The feature is therefore
often referred to as a \textsl{Cyclotron Resonant Scattering Feature} (CRSF).  
The energy spacing between the Landau levels is given by E$_{\rm cyc}$ =
$\hbar$eB/m$_{\rm e}$c = 11.6\,\text{keV}\,B$_{12}$, where
B$_{12}$=B/10$^{12}$\,\text{G}, providing a direct method of measuring
the magnetic field strength at the site of the emission of the X-ray spectrum. The 
observed line energy is subject to gravitational redshift $z$ at the location where 
the line is formed, such that the magnetic field may be estimated by
B$_{12}$ = (1+z)~E$_\text{obs}$/11.6\,${\rm keV}$, with E$_\text{obs}$ being
the observed cyclotron line energy.  The discovery of the
cyclotron feature in the spectrum of Her X-1 provided the first ever
direct measurement of the magnetic field strength of a neutron star,
in the sense that no other model assumptions are needed.  
Originally considered an exception, cyclotron features are now known to 
be rather common in accreting X-ray pulsars, with $\sim20$ binary pulsars 
now being confirmed cyclotron line sources, with several objects showing 
multiple lines (up to four harmonics in 4U~0115+63). Reviews are given by 
e.g., \citet{Coburn_etal02,Staubert_03,Heindl_etal04,Terada_etal07,Wilms_12,
CaballeroWilms_12}. Theoretical calculations of  cyclotron line spectra
have been performed either analytically \citep{Ventura_etal79,Nagel_81,Nishimura_08}
or making use of Monte Carlo techniques 
\citep{ArayaHarding_99,ArayaGochezHarding_00,Schoenherr_etal07}.

Here we present new results (from the last five
years) on the energy of the cyclotron resonance scattering feature E$_{\rm cyc}$
in the pulse averaged X-ray spectrum of Her~X-1, combined with the 
historical long-term evolution. We present the first statistically significant 
evidence of a true long-term decrease in E$_{\rm cyc}$ and first evidence of a weak
dependence of E$_{\rm cyc}$ on phase of the 35\,d cycle. We speculate about
the physics behind both the long-term decrease and the previously
observed fast upward jump as being connected to changes in the 
configuration of the magnetic field structure at the responsible polar cap 
of the accreting neutron star. A preliminary report was published earlier 
by \citet{Staubert_13}.

\section{Data base and method of analysis}
\label{sec:data_base}

Her~X-1 is probably the best observed accreting binary X-ray pulsar.
Its X-ray spectrum, including the CRSF, has been measured by many
instruments since the discovery of the CRSF in 1976 \citep{Truemper_etal78}. 
The data base behind previously reported results are summarized in
corresponding Tables of the following publications: \citet{Gruber_etal01,Coburn_etal02,
Staubert_etal07,Staubert_etal09,Staubert_etal13,Klochkov_etal08,Klochkov_etal11,
Vasco_etal11,Vasco_etal13}.
Details about more recent observations (proposed by our group during the last five years) 
by \textsl{RXTE}, \textsl{INTEGRAL}, \textsl{Suzaku} and \textsl{NuSTAR}
are given in Tables \ref{tab:new_obs} and \ref{tab:new_results}.
Of particular importance is the observation by \textsl{NuSTAR} in September 2012
which provided the most accurate value for the CRSF energy measured to date
\citep{Fuerst_etal13}.
For the investigation of the long-term evolution of the cyclotron line energy,
\textsl{Main-On} state observations at 35-day-phases $<0.20$ were used,
in order to avoid interference with a dependence on 35\,d phase (see
Sec. \ref{sec:35d_phase}).
The spectral analysis was performed using the standard software appropriate 
for the respective satellites (see the publications cited above), with the addition
of non-standard corrections based on our deeper analysis of the calibration of IBIS
(Imager on Board the INTEGRAL Satellite, \citealt{Ubertini_etal03}),
and the use of RECORN models \citep{Rothschild_etal11} for \textsl{RXTE}.
For the spectral model we have chosen the \texttt{highecut}\footnote{
http://heasarc.nasa.gov/xanadu/xspec/manual/XSmodelHighecut.html}
model which is based on a power law continuum with exponential cut-off, and 
the CRSF is modeled by a multiplicative absorption line with a Gaussian 
optical depth profile. 
Details of the fitting procedure can be found in the papers cited above. 

\begin{table}
\caption[]{Details of recent observations of Her~X-1 by \textsl{INTEGRAL}, \textsl{RXTE}, 
\textsl{Suzaku} and \textsl{NuSTAR}.}
    \label{tab:new_obs}
\vspace{-3mm}
\begin{center}
\begin{tabular}{lllll}
\hline\noalign{\smallskip}
Observatory          & Date of          & Center     & Obs ID             & Expos.  \\
                              & observation            & MJD         &                         & [ksec]    \\
\hline\noalign{\smallskip}
\textsl{INTEGRAL}& 2007 Sep 03-08  & 54348.0  &  Rev. 597/598    & 414.72  \\      
\textsl{RXTE}        & 2009 Feb 04-05   & 54866.6  &  P 80015            & ~~22.19   \\    
\textsl{INTEGRAL}& 2010 July 10-18  & 55390.0  &  Rev. 945-947    & 621.03  \\       
\textsl{Suzaku}      & 2010 Sep 22       & 55462.0  &  405058010/20   &  ~~41.66   \\   
\textsl{Suzaku}      & 2010 Sep 29       & 55468.0  &  405058030/40   &  ~~45.66   \\   
\textsl{INTEGRAL}& 2011 June 25-27 & 55738.3  &  Rev. 1062         & 95.9       \\       
\textsl{INTEGRAL}& 2011 July 03-05  & 55744.5   &  Rev. 1069         & 107.8       \\       
\textsl{INTEGRAL}& 2012 April 1-4     &  56019.5  & Rev. 1156          & 42.6       \\       
\textsl{Suzaku}      & 2012 Sep 19-25  & 56192.2  &  4070510-          &  $\sim 70$   \\  
                              &                            &                &  10, 20, 30          &  \\
\textsl{NuSTAR}     & 2012 Sep 19-25  & 56192.2  &  3000200600-    &  ~~72.9   \\      
                              &                             &                &  2, 3, 5, 7            &  \\
\noalign{\smallskip}\hline
   \label{tab:new_obs}
\end{tabular}\end{center}
\end{table}

\section{Variation of the cyclotron line energy E$_{\rm cyc}$}

Variability in the energy of the CRSF in Her X-1  
is found with respect to the following variables: \\
- Variation with X-ray luminosity (both on long and on short \\
\indent timescales). \\ 
- Variation with phase of the 1.24\,s pulsation. \\
- Variation with phase of the 35\,d precessional period. \\
- Variation with time, that is a true long-term decay. \\
The dependence on pulse phase, which is described in detail by \citet{Vasco_etal13},
will only be discussed here to the extent necessary to understand the connection to the 
35\,d phase dependence.

\begin{figure}
\includegraphics[angle=90,width=10.5cm]{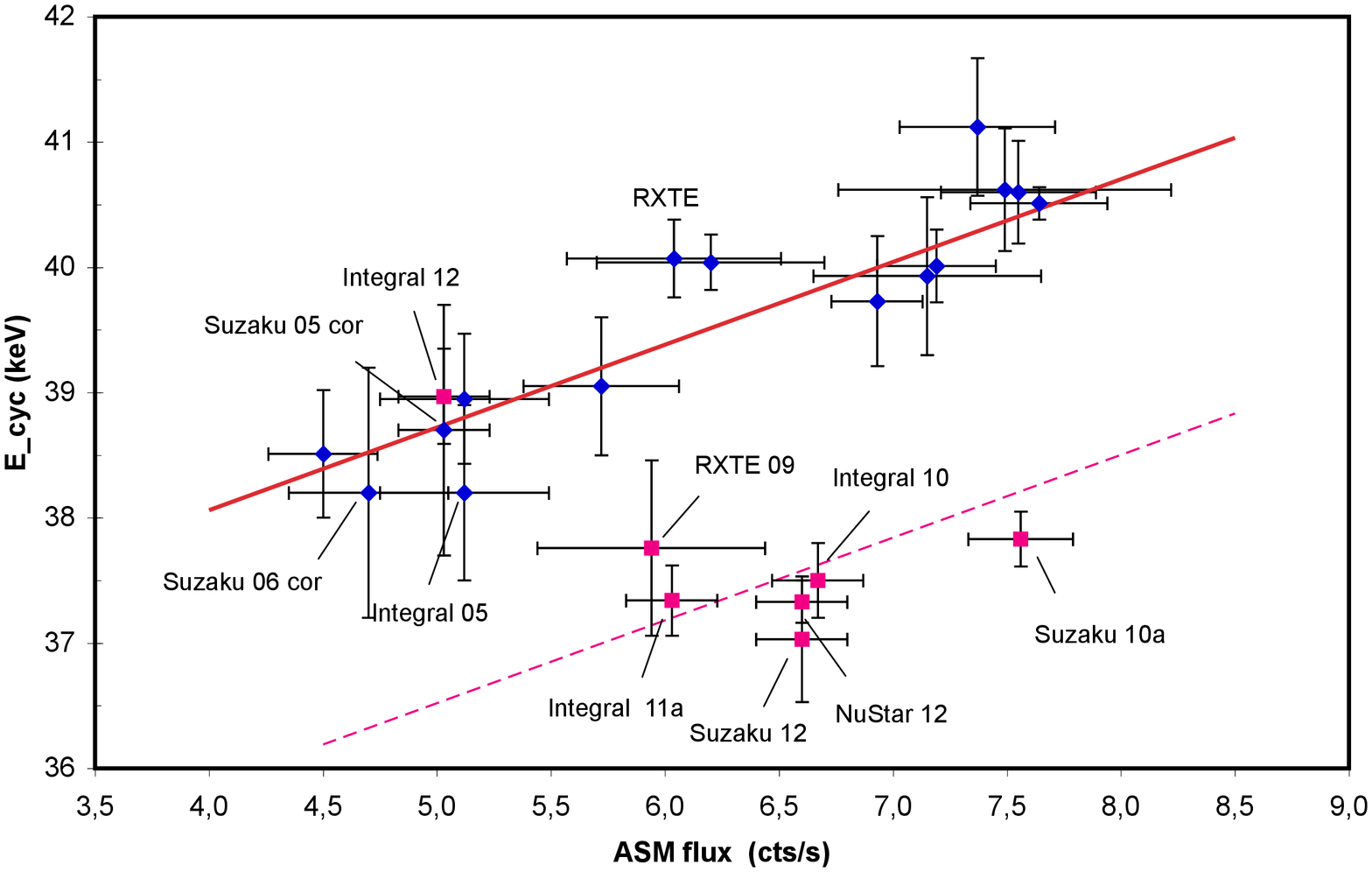}
\hfill
\vspace{-1.3cm}
\caption{The positive correlation between the cyclotron line energy 
  and the maximum X-ray flux of the corresponding 35-day cycle as  
  measured by \textsl{RXTE/ASM} (see Fig.~2 of \citet{Staubert_etal07}) with eight added points:
  \textsl{INTEGRAL} 2005 \citep{Klochkov_etal08}, \textsl{Suzaku} of 2005 and 2006 \citep{Enoto_etal08}, 
  \textsl{RXTE} 2009, \textsl{INTEGRAL} 2010, \textsl{Suzaku} 2010 and 2012 and \textsl{NuSTAR} 2012.
  The \textsl{Suzaku} points of 2005/2006 have been corrected upward by 2.8\,keV, 
  to account for the difference arising because the Lorentzian profile was used in the analysis by  
  \citet{Enoto_etal08}, while for all others the Gaussian profile was used. 
  The blue rhombs are values observed until 2006, the red dots are from after 2006.
  The solid red line is a linear fit to data until 2006 with the original slope of
  0.66\,keV/(ASM cts/s), as found by \citet{Staubert_etal07}.
  The dotted red line is the best fit to the data after 2006 with the slope fixed to the same value.}
   \label{fig:correlation}
\end{figure}

\section{Variation of E$_{\rm cyc}$ with luminosity}
\label{sec:luminosity-dependence}

For Her~X-1, the dependence of the centroid energy of the phase averaged cyclotron 
line on X-ray flux was discovered by \citet{Staubert_etal07} while analyzing a uniform
set of observations from \textsl{RXTE}. The original aim of the analysis at that time had 
been to investigate a possible decrease in the phase averaged cyclotron line energy with 
time during the first decade of \textsl{RXTE} observations. Instead, the dependence on 
X-ray flux was discovered and shown that the apparent decrease in the measured
values of the line energy (see Section~\ref{sec:secular})
was largely an artifact due to this flux dependence. 
The correlation was found to be positive, that is the cyclotron line energy E$_{\rm cyc}$ 
increases with increasing X-ray luminosity $L_x$. 

Fig.~\ref{fig:correlation} reproduces the original correlation graph of \citet{Staubert_etal07}
with new data points added (see Table~\ref{tab:new_results}). The first three new data points
(INTEGRAL~05 and Suzaku~05/06) fit very well into the previous data set (and do not change
the formal correlation - see the solid red line), but most of the values from 2006-2012 are significantly lower. 
As we will show below, it is these data which clearly establish a decrease in the cyclotron line energy 
with time.
After 2006 the flux dependence is less obvious. However, the data points (except the one from \textsl{INTEGRAL} 2012)
are consistent with the originally measured slope (0.66 keV/ASM-cts/s) with generally lower E$_{\rm cyc}$ values. 
The dotted red line is a fit through the data after 2006 with the same slope as the solid red line.
We note that \textsl{flux} refers to the maximum \textsl{Main-On} flux as determined using the \textsl{RXTE}/ASM
and/or the \textsl{Swift}/BAT monitoring data (since 2012 from BAT only); the conversion is: 
(2-10\,keV ASM-cts/s) = 89 $\times$ (15-50\,keV BAT-cts~cm$^{-2}$~s$^{-1}$).
The \textsl{INTEGRAL} 2012 point does clearly not follow this behavior, as will be more obvious below. 
We have invested a considerable effort to check the calibration of the \textsl{INTEGRAL}/ISGRI detector 
(INTEGRAL Soft Gamma-Ray Imager, \citealt{Lebrun_etal03}) for the time of observation and the 
data analysis procedure. The ISGRI response was closely examined by us for each of our Her X-1 observations. 
When necessary, the ARFs (Auxilliary Response Files) were checked (using the nearest Crab observations) and the 
energy scale was individually controlled by making use of observed instrumental background lines 
with known energy. Finally, spectra were generated using data from 
SPI (Spectrometer onboard INTEGRAL, \citealt{Vedrenne_etal03}): the resulting E$_{\rm cyc}$ 
values were always consistent with those of the ISGRI analysis.
Since we have found no errors, we keep this point in our data base, but will exclude it from some of the 
analysis discussed below.

\section{Variation of E$_{\rm cyc}$ with precessional phase}
\label{sec:35d_phase}

In order to investigate whether the cyclotron line energy has any dependence on phase
of the 35\,d precession, we had successfully scheduled a few \textsl{Main-On} observations at 
late 35\,d phases. In addition to the full coverage of the \textsl{Main-On} of cycle no. 232 
(a singular event!), we so far have four more measurements: 
from \textsl{INTEGRAL} in July/August 2007 and July 2011, as well as  from \textsl{Suzaku} in 
September 2011 (all at 35\,d phases 0.25, see Table~\ref{tab:recent_values}) and from 
\textsl{INTEGRAL} in July 2005 at phase 0.24 (E$_{\rm cyc}$ = $(37.3\pm1.2)$\,keV and 
$(5.12\pm0.37)$\,ASM-cts/s). As we will show below, the cyclotron line energy changes with
time. So, for a comparison of values measured at different times, they must be normalized
to a common reference time. We are therefore making use of the results presented in
Section \ref{sec:secular} and compare E$_{\rm cyc}$ values which are normalized to the
reference time of MJD~53500 (using the slope of fit~4 of Table~\ref{tab:3D_3}):
($37.11\pm0.61$)\,keV, ($40.15\pm0.81$)\,keV, ($36.92\pm1.09$)\,keV, and 
($37.36\pm1.20$)\,keV, respectively. Three out of these four values are indeed 
quite low in comparison to all other time normalized values, and gives an indication that 
E$_{\rm cyc}$ may indeed decrease at late 35\,d phases. However, because of the lower fluxes 
at late 35\,d phases, the uncertainties are fairly large for all of these measurements.

\begin{figure}
 \includegraphics[width=0.51\textwidth]{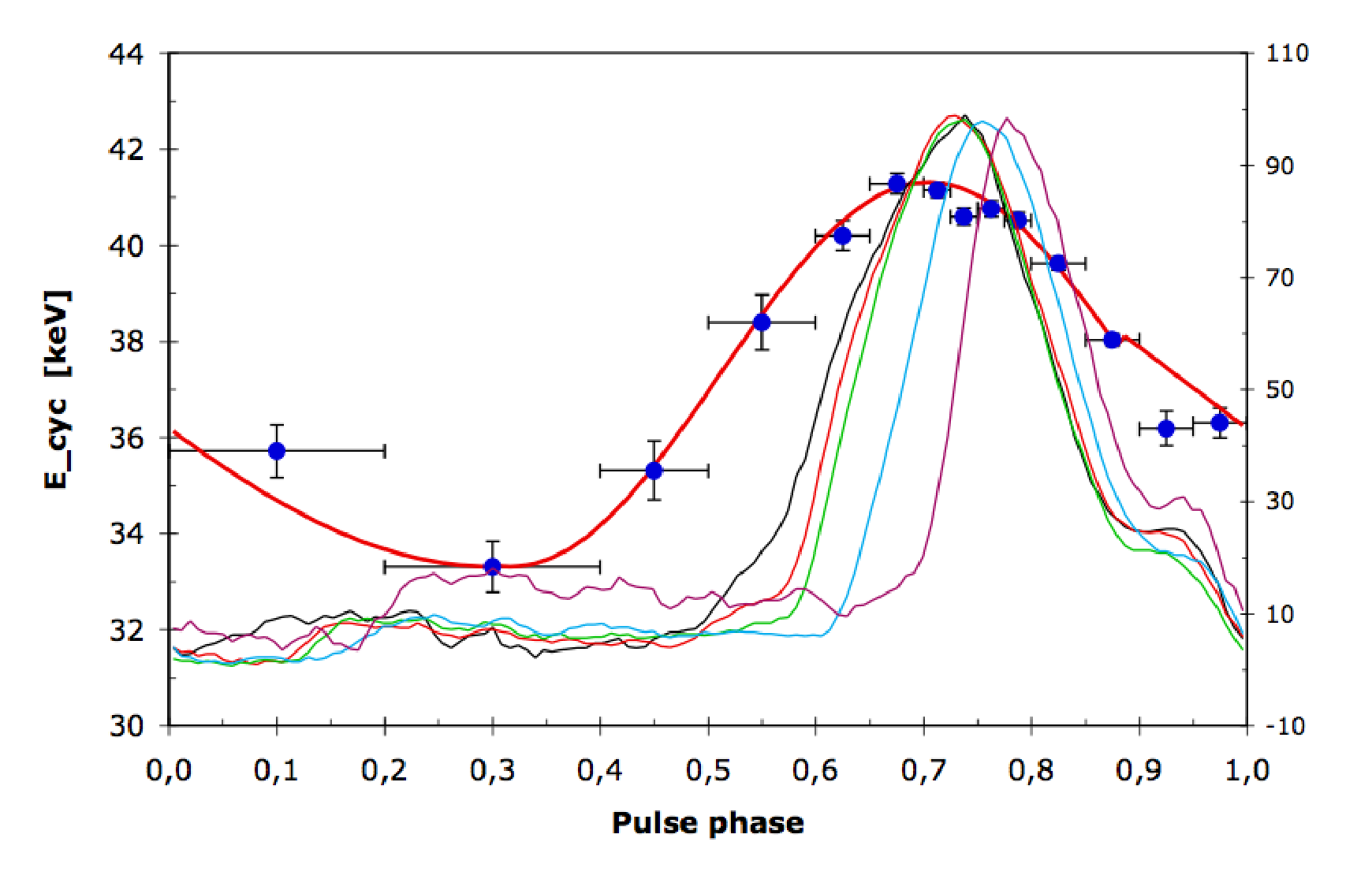}    
\hfill
\vspace{-4mm}
\caption{Mean dependence of cyclotron line energy on pulse phase for the \textsl{Main-On} of 35\,d 
       cycle 323, as observed by \textsl{RXTE}/PCA in 2002 November. The solid red line represents a best fit function
       (a combination of two cosine components). Normalized pulse profiles of the 30-45\,keV range are shown for five
       different 35\,d phases: 0.048 (black), 0.116 (red), 0.166 (green), 0.21 (blue) and 0.24 (purple). The main pulse is 
       progressively moving to the right. The right hand scale is normalized flux (0-100) for the pulse profiles.}
   \label{fig:pulse_phase}
\end{figure}

A more indirect, but perhaps more reliable method is the following:
There are two well established observational facts with regard to pulse profiles and cyclotron 
line energies in Her~X-1, both demonstrated in Fig.~\ref{fig:pulse_phase}: \\
\vspace{-5mm}
\begin{enumerate}
\item E$_{\rm cyc}$ varies strongly with pulse phase (by up to $\sim25$\%) 
\citep{Voges_etal82,Soong_etal90, Vasco_etal13}. The shape of the 
(E$_{\rm cyc}$ vs pulse phase)-profile, is not dependent on 35\,d phase \citep{Vasco_etal13}.
\item The main peak of the pulse profile moves to later pulse phases with increasing 35\,d phase  
\citep{Staubert_etal09}. This is also true for the 30-45\,keV profiles -- the energy range which 
includes the CRSF (these profiles are shown in Fig.~\ref{fig:pulse_phase}).
\end{enumerate}

The combination of these two observational facts inevitably leads to a modulation of E$_{\rm cyc}$ 
with 35\,d phase: with progressing 35\,d phase, more and more photons are found at later pulse 
phases (in the main peak of the pulse) where the cyclotron line energy is decreasing. This means 
that the phase averaged cyclotron line energy must decrease with progressing 35\,d phase
(the above mentioned effect was first considered by \citet{Klochkov_etal08b}). 
In order to quantitatively test this, we have performed a formal folding of 30-45\,keV pulse profiles 
(for 10 different 35\,d phases) with template (E$_{\rm cyc}$ vs pulse phase) profiles for these 
35\,d phases. The template profiles were constructed by inter- and extrapolation of the four 
individual profiles at different 35\,d phases as given in Fig. 3 and 4 of \citet{Vasco_etal13} 
(taking into account that both the maximum value and the peak-to-peak amplitude are slightly 
35\,d phase dependent). By folding the pulse profile with the corresponding E$_{\rm cyc}$ profile, 
the expected pulse phase averaged E$_{\rm cyc}$ value can be calculated. 

\begin{figure}
\includegraphics[width=0.47\textwidth]{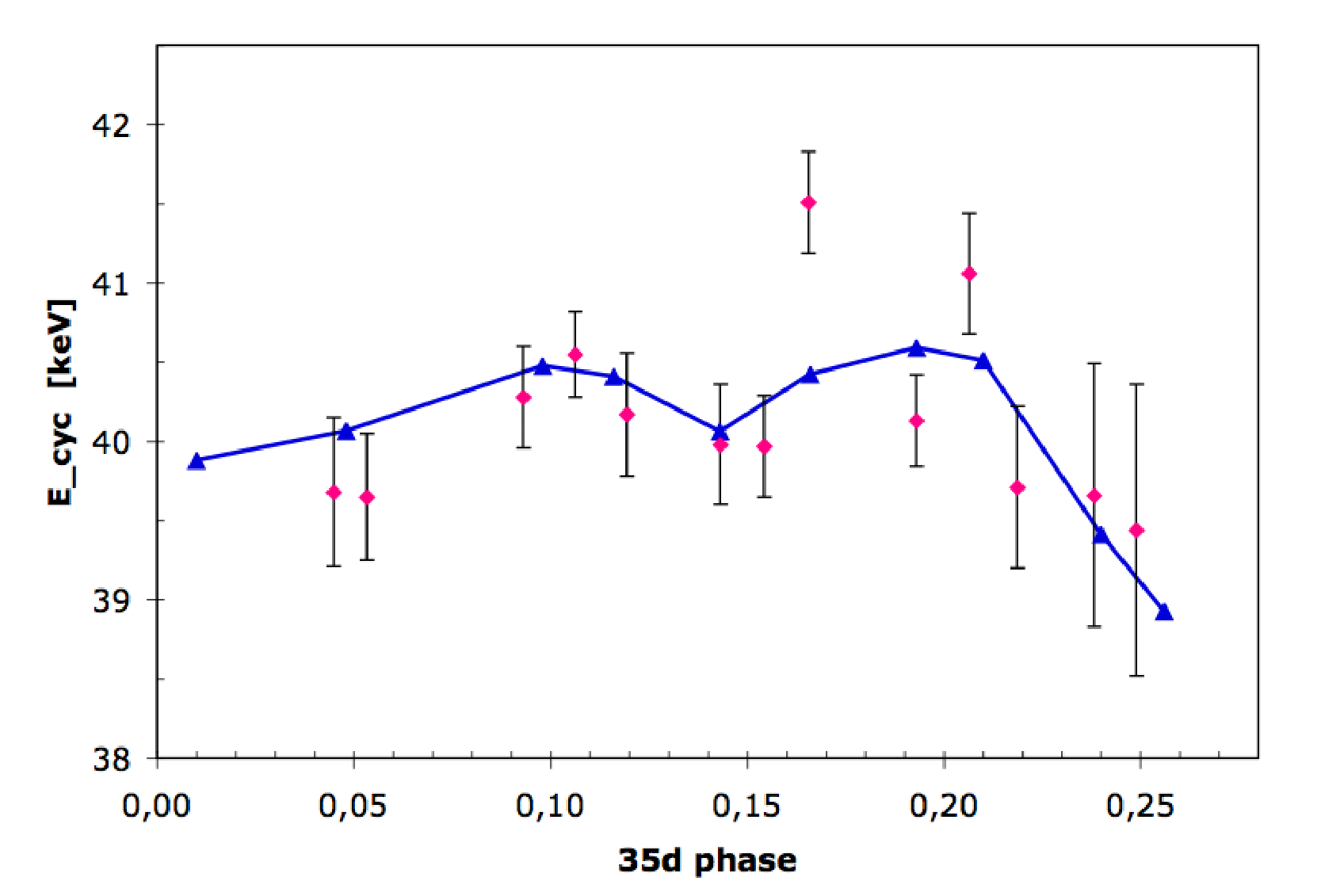} 
\hfill
\caption{Centroid pulse phase averaged cyclotron line energies at different 35\,d phases of 
    \textsl{Main-On} cycle 323. The data points with uncertainties are direct measurements for ten small integration 
    intervals. The blue triangles connected by the solid blue line are values which are calculated by folding observed 
    pulse profiles in the 30-45\,keV range with template (E$_{\rm cyc}$ vs pulse phase)-profiles for the same 35\,d phases
    (see text for a detailed description). }
   \label{fig:35d}
\end{figure}

In Fig.~\ref{fig:35d} these calculated values are shown as blue triangles (connected by the solid 
blue line): a slow increase up to phase $\sim0.19$ is followed by a somewhat sharper decay.
For comparison, we show the directly measured phase averaged E$_{\rm cyc}$ values (data 
points with uncertainties) for 13 small integration intervals covering the \textsl{Main-On} of 
cycle 323 (MJD 52599/Nov 2002). The directly measured values have relatively large uncertainties, 
but are overall consistent with the calculated modulation, and also with regard to the mean absolute value.

\begin{table*}
\caption[]{Recent cyclotron line energy measurements during \textsl{Main-Ons} 
of Her~X-1 by \textsl{INTEGRAL}, \textsl{RXTE}, \textsl{Suzaku} and \textsl{NuSTAR}. 
Uncertainties are at the 68\% level. 35\,d cycle numbering and 35\,d phase is according 
to \citet{Staubert_etal83,Staubert_etal13}. \\
}
       \label{tab:new_results}
\begin{center}
\vspace{-2mm}
\begin{tabular}{lllllllll}
\hline\noalign{\smallskip}
Satellite & Observation & 35\,d & Center     & 35\,d   & observed                    & max. Flux              & Flux normalized         & References\\
              & year/month  & cycle & [MJD]      &  phase & E$_{\rm cyc}$ [keV]   & [ASM cts/s]$^{1}$  & E$_{\rm cyc}$ [keV]$^{3}$   &\\
\hline\noalign{\smallskip}
\textbf{\textsl{RXTE}}    & 2009 Feb        & 388  &  54866.60  & 0.11 & $37.76\pm0.70$   &  $5.94\pm0.50$ & $38.14\pm0.71$ & this work\\ 
\textbf{\textsl{Suzaku}} & 2005 Oct         & 353  &  53648.00  & 0.11 & $38.70\pm1.00$   &  $5.03\pm0.20$ & $39.47\pm1.02$ & \citet{Enoto_etal08}$^{2}$ \\
                                     & 2006 Mar        & 358  &  53824.20  & 0.19 & $38.20\pm1.00$   &  $4.70\pm0.35$ & $39.12\pm1.02$ & \citet{Enoto_etal08}$^{2}$ \\
                                     & 2010 Sep        & 405  &  55462.00  & 0.07 & $37.83\pm0.22$   &  $7.56\pm0.23$ & $37.50\pm0.22$ & this work \\ 
                                     & 2010 Sep        & 405  &  55468.00  & 0.25 & $36.50\pm1.05$   &  $7.56\pm0.23$ & $35.17\pm1.04$ & this work \\ 
                                     & 2012 Sep        & 426  &  56192.23  & 0.09 & $37.03\pm0.50$   & $6.60\pm0.50$  & $37.12\pm0.50$ & this work \\
\textbf{\textsl{INTEGRAL}} & 2010 July   & 403 & 55390.00   & 0.05 & $37.50\pm0.30$    & $6.67\pm0.20$  & $37.56\pm0.30$ & this work \\ 
 & 2011 June       & 413 & 55738.26   & 0.08 & $37.34\pm0.28$   & $6.03\pm0.20$ & $37.68\pm0.28$ & this work \\ 
 & 2007 July/Aug & 373 & 54348.00   & 0.25 & $36.50\pm0.60$   & $4.50\pm0.48$ & $37.51\pm0.62$ & this work \\ 
 & 2011 July       & 413 & 55744.50   & 0.25  & $38.53\pm0.78$  & $7.00\pm0.20$ & $38.44\pm0.78$ & this work \\  
 & 2012 Aug       & 421 & 56019.50   & 0.09  & $38.97\pm0.38$   & $5.03\pm0.20$ & $39.74\pm0.39$ & this work \\  
\textbf{\textsl{NuSTAR}} & 2012 Sep         & 426  &  56192.20  & 0.09 & $37.33\pm0.17$  & $6.60\pm0.20$  & $37.42\pm0.17$ & \cite{Fuerst_etal13} \\ 
\noalign{\smallskip}\hline
   \label{tab:recent_values}
\end{tabular}
\end{center}
\vspace{-1mm}
$^{1}$ The maximum \textsl{Main-On} flux was determined using the monitoring data of \textsl{RXTE}/ASM
and \textsl{Swift}/BAT (since 2012 from BAT only); the conversion is: (2-10\,keV ASM-cts/s) = 89 $\times$ 
(15-50\,keV BAT-cts~cm$^{-2}$~s$^{-1}$) \\
$^{2}$ The two \textsl{Suzaku} data points from 2005 and 2006 are from \citet{Enoto_etal08},
adjusted to describing the cyclotron line by a Gaussian line profile (see text). \\
$^{3}$ These values refer to this work: normalization to an ASM flux of 6.8 (ASM-cts/s) using the E$_{\rm cyc}$/flux relationship 
(slope of 0.44\,keV/ASM-cts/s) of the two-variable fit to the 1996-2012 data set (see Table~\ref{tab:3D_3}).
\end{table*}

\section{Variation of E$_{\rm cyc}$ with time - long-term variation}
\label{sec:secular}

\begin{figure*}
\includegraphics[width=0.65\textwidth,angle=-90]{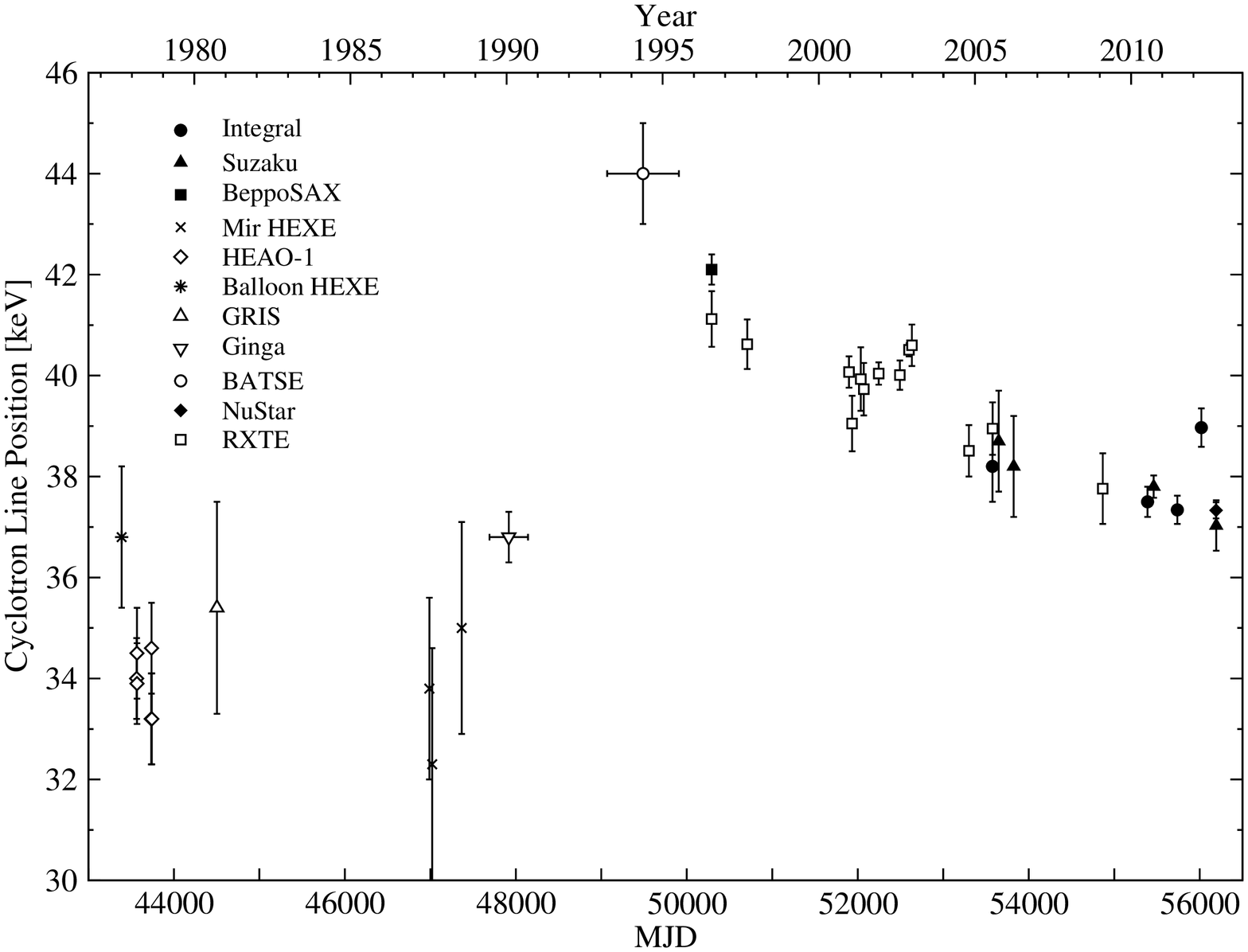}
\vspace{-8mm}
\caption{The centroid energy of the phase averaged cyclotron resonance
  line feature in Her~X-1 since its discovery. Data from before 1997
  were originally compiled by \citet{Gruber_etal01}, where the original 
  references can be found. 
The data after 1997 are from observations by \textsl{RXTE}, \textsl{INTEGRAL} 
\citep{Klochkov_etal06,Staubert_etal07, Klochkov_etal08} and \textsl{Suzaku}
\citep{Enoto_etal08}, plus  recent values as given in Table~\ref{tab:recent_values}.
Here only values measured at 35d phases $<$0.20 are shown.
}
 \label{fig:history}
\end{figure*}

In Fig.~\ref{fig:history} (an update of Fig.~\ref{fig:correlation} of \citealt{Staubert_etal07}) we display 
observed values of the pulse phase averaged centroid cyclotron line energy as a 
function of time, covering the complete history of observations since the discovery of 
the line in 1976. We combine historical data, as taken from the compilation by 
\citet[their Tables~2 and~3]{Gruber_etal01} for the time before the \textsl{RXTE} era, 
published values from observations with \textsl{RXTE} and \textsl{INTEGRAL} 
\citep{Klochkov_etal06,Staubert_etal07, Klochkov_etal08} and with
\textsl{Suzaku} \citep{Enoto_etal08}, as well as recent values as given in 
Table~\ref{tab:recent_values} (see also \citealt{Staubert_13,Fuerst_etal13}).
For the analysis of the long-term variation of E$_{\rm cyc}$ we exclude values with 
35\,d phases $>$0.20 in order to avoid contamination due to a possible third variable, 
the 35\,d phase (a dependence, if any, is very weak for small phases; see Fig.~\ref{fig:35d}).

Two features are apparent from Fig.~\ref{fig:history}:
Firstly, we confirm the apparent difference in the mean cyclotron line energy 
before and after 1991, first pointed out by \citet{Gruber_etal01}.
Taking the measured values of E$_{\rm cyc}$ and their stated uncertainties
at face value, the mean cyclotron line energies $\langle E_c \rangle$ 
from all measurements before 1991 is $34.9\pm0.3\,\text{keV}$, the corresponding 
value for all measurements between 1991 and 2006 is $40.3\pm0.1\,\text{keV}$ 
($40.2\pm0.1\,\text{keV}$ for \textsl{RXTE} results only, showing that
the very high value measured by \textsl{BATSE} is not decisive). However, a 
comparison of measurements from different instruments is difficult because of 
systematic uncertainties due to calibration and analysis techniques. 
Nevertheless, we believe that the large difference of $\sim 5\,\text{keV}$ 
between the mean values and the good internal consistency within the
two groups (5 different instruments before 1991 and four after 1991) most 
likely indicate real physics.

As already mentioned in Sec.~\ref{sec:luminosity-dependence}, 
the first observations with \textsl{RXTE} in 1996 and 1997 showed lower 
E$_{\rm cyc}$ values than those found from \textsl{CGRO}/BATSE and 
\textsl{Beppo}/SAX, leading to the idea of a possible long-term decay.
This idea had then served successfully as an important argument to ask for 
more observations of Her~X-1. In a series of \textsl{RXTE} observations until 
2005 the apparent decrease seemed to continue until this date. At this time
we were determined to publish a paper claiming evidence of a decay 
of the phase averaged cyclotron line energy E$_{\rm cyc}$.
However, working with a uniform set of \textsl{RXTE} data between
1996 and 2005, we discovered that there was a dependence of E$_{\rm cyc}$ 
on X-ray flux \citep{Staubert_etal07}, degrading the apparent decrease with 
time largely to an artifact: nature seemed to have conspired such that later 
measurements were (on average) taken when the flux happened to be low 
(Her X-1 is known for varying its flux within a factor of two, on timescales of a 
few 35\,d cycles). When the cyclotron line energy was normalized to a common 
flux value, the time dependence 
vanished.

In the following we will demonstrate in a systematic way that today we have clear evidence 
for a reduction in the phase averaged cyclotron line energy E$_{\rm cyc}$
with time over the last 20 years. Both dependencies - on flux and on time - seem to 
be always present (they  may, however, change their relative importance with time). 
Using a procedure of fitting with two variables simultaneously, the 
two dependencies can be separated and the formal correlation minimized.
With the inclusion of new measurements (2005-2012), we are now able to present
the first statistically significant evidence of a \textsl{true long-term decay of 
the phase averaged cyclotron line energy}. 

\begin{figure*}
\includegraphics[width=0.6\textwidth,angle=90]{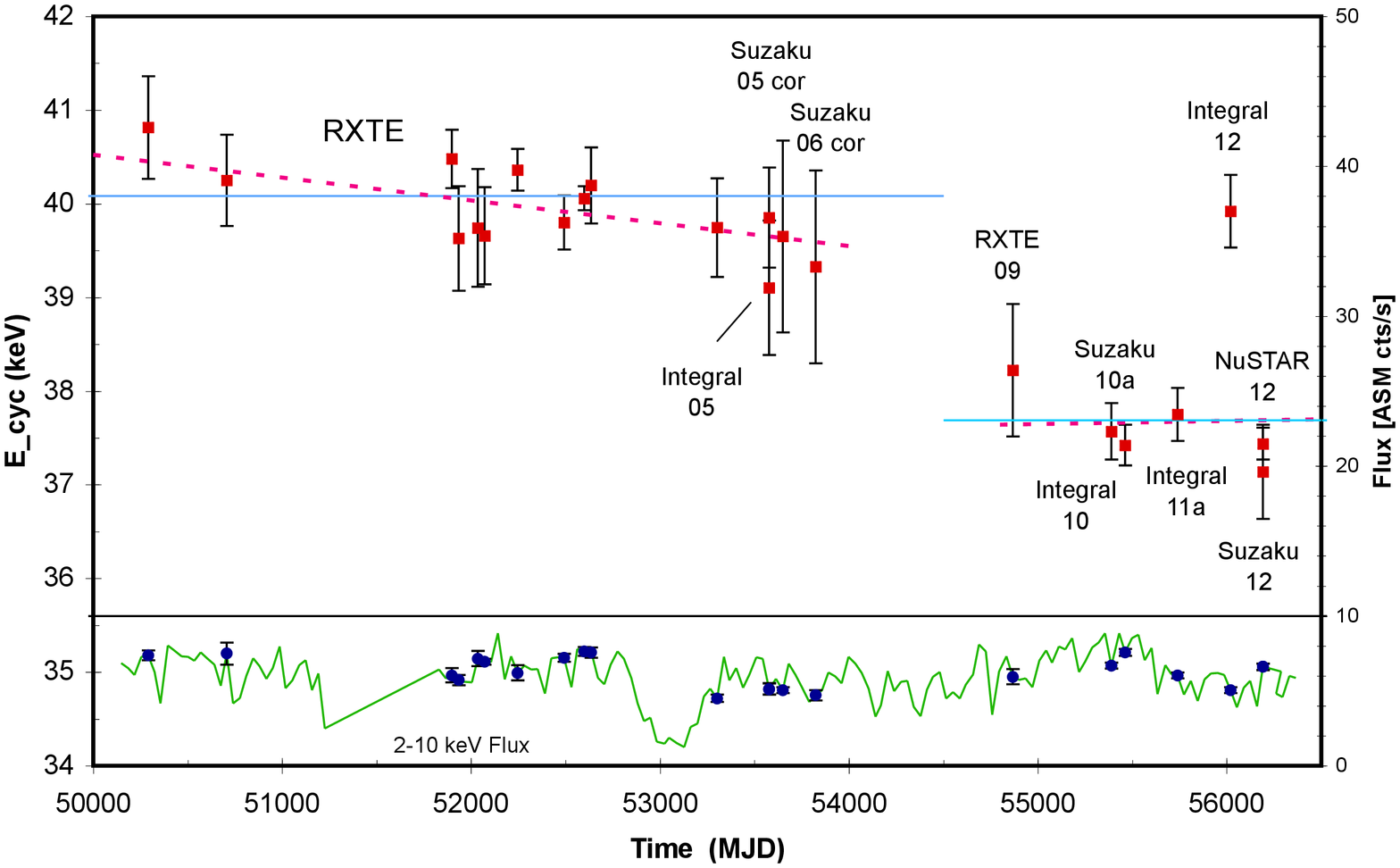}
\hfill
\vspace{-1cm}
\caption{\textsl{Upper panel, left scale:} Her~X-1 pulse phase averaged cyclotron line energies E$_{\rm cyc}$ 
  normalized to a reference ASM count rate of 6.8\,cts/s using a flux dependence of 0.54\,keV/ASM-cts/s. 
  A break in mean E$_{\rm cyc}$ after MJD 54000 ($>$ 2006) is apparent.
  \textsl{Lower panel, right scale:} 2-10 keV X-ray flux (of 35\,d maximum) from monitoring by \textsl{RXTE}/ASM
  (from \textsl{Swift}/BAT after MJD~56000). Blue data points with uncertainties are those fluxes which
  are used to correlate with E$_{\rm cyc}$, the green curve connects measurements of each 35\,d
  cycle.
 }
 \label{fig:Fnorm2}
\end{figure*}

\subsection{Normalizing E$_{\rm cyc}$ using the originally discovered flux dependence}
\label{sec:flux_norm1}

Before introducing fits with two variables (flux and time), we repeat here the procedure
applied in \citet{Staubert_13}, that neglects any time dependence and normalizes 
E$_{\rm cyc}$ by flux only, using the originally determined linear dependence with a
slope of 0.66 keV/(ASM-cts/s) \citep{Staubert_etal07}. The normalization is done to the common 
reference flux of 6.80 ASM-cts/s. The results were shown in Fig.~2(right) of \citet{Staubert_13} 
(and, slightly updated, in Fig.~13 of \citet{Fuerst_etal13}), showing the flux normalized cyclotron 
line energy as a function of time (1996-2012). These figures demonstrate that there is a significant
decrease in E$_{\rm cyc}$ at least after 2006 ($>$MJD 54500). Based on a repeated
analysis we give the following additional information. \\
(1) The normalized E$_{\rm cyc}$ values between 1996 and 2006 (MJD 50000-54000) 
are consistent with a constant (based on $\chi^{2}$/dof = 0.77), supporting the neglect of a 
time dependence for this decade. The weighted mean is ($40.06\pm0.09$) keV.
However, even during this time period there is a slight downward trend with a slope of 
$(-2.5\pm1.4)\times{10^{-4}}\,\rm keV d^{-1}$. In order to test whether this downward 
trend is really a true time dependence and not an artifact due to the neglect of e.g., a more 
complicated flux dependence, a quadratic term was added when doing the flux normalization: 
this does not improve the fit and does not remove the downward trend.\\
(2) The weighted mean of the normalized E$_{\rm cyc}$ values for 2006-2012 is
($37.69\pm0.10$)\,keV (with a high $\chi^{2}$/dof of 6.6, mainly due to the \textsl{INTEGRAL} 
2012 point). \\
(3) The difference between these two mean values is highly significant ($>$17 standard 
deviations), demonstrating the decrease in E$_{\rm cyc}$ with time.

\begin{table}
\caption[]{Details of fits with formula (1) to E$_{\rm cyc}$ values observed between
1996 and 2006. The reference time is T$_{0}$ = 53000.}
\vspace{-3mm}
\begin{center}
\begin{tabular}{llllll}
\hline\noalign{\smallskip}
Param.                & E$_{\rm 0}$        & a [keV/                  & b [$10^{-4}$         & $\chi^{2}$  & dof  \\
fitted                    & [keV]                   &ASM-cts/s]           & keV/d]                  &           & \\
\hline\noalign{\smallskip}
E$_{\rm 0}$         & $40.15\pm0.09$  & 0.00                    &  0.00                     & 47.6   &  14 \\    
E$_{\rm 0}$, a     & $40.08\pm0.09$  & $0.58\pm0.10$   &  0.00                     & 10.9   &  13 \\  
E$_{\rm 0}$, b     & $39.88\pm0.12$  & 0.00                    &  $-4.60\pm1.43$   & 37.3   &  13 \\   
E$_{\rm 0}$, a, b & $39.88\pm0.12$  & $0.54\pm0.10$   &  $-2.91\pm1.47$   &   7.0   &  12 \\   
\noalign{\smallskip}\hline
   \label{tab:3D_2}
\end{tabular}\end{center}
\end{table}

\subsection{Normalizing E$_{\rm cyc}$ using fits to the 1996-2006 data with two variables}
\label{sec:flux-time_norm2}

From Fig.~\ref{fig:correlation}, it is already evident that most of the E$_{\rm cyc}$ 
values measured after 2006 are significantly lower than those from before.
In order to separate the dependence on time and the dependence on X-ray flux, 
we have performed fits to the 1996-2006 data with two variables - X-ray flux and
time. We use the function 
\begin{equation}
{\rm E}_{\rm cyc}(\rm calc) = E_{\rm 0} + a \times (F - F_{0}) + b \times (T - T_{0}) \\
\end{equation}
with F being the X-ray flux (the maximum flux of the respective 35\,d cycle) in units of 
ASM-cts/s, as observed by \textsl{RXTE}/ASM (and/or \textsl{Swift}/BAT), with  
F$_{0}$ = 6.80 ASM-cts/s, and T being time in MJD with T$_{0}$ = 53000
(the relationship between ASM and BAT is the folllowing: (2-10\,keV ASM-cts/s) = 89 * 
(15-50\,keV BAT-cts~cm$^{-2}$~s$^{-1}$). 

\begin{figure*}
\includegraphics[width=0.6\textwidth,angle=90]{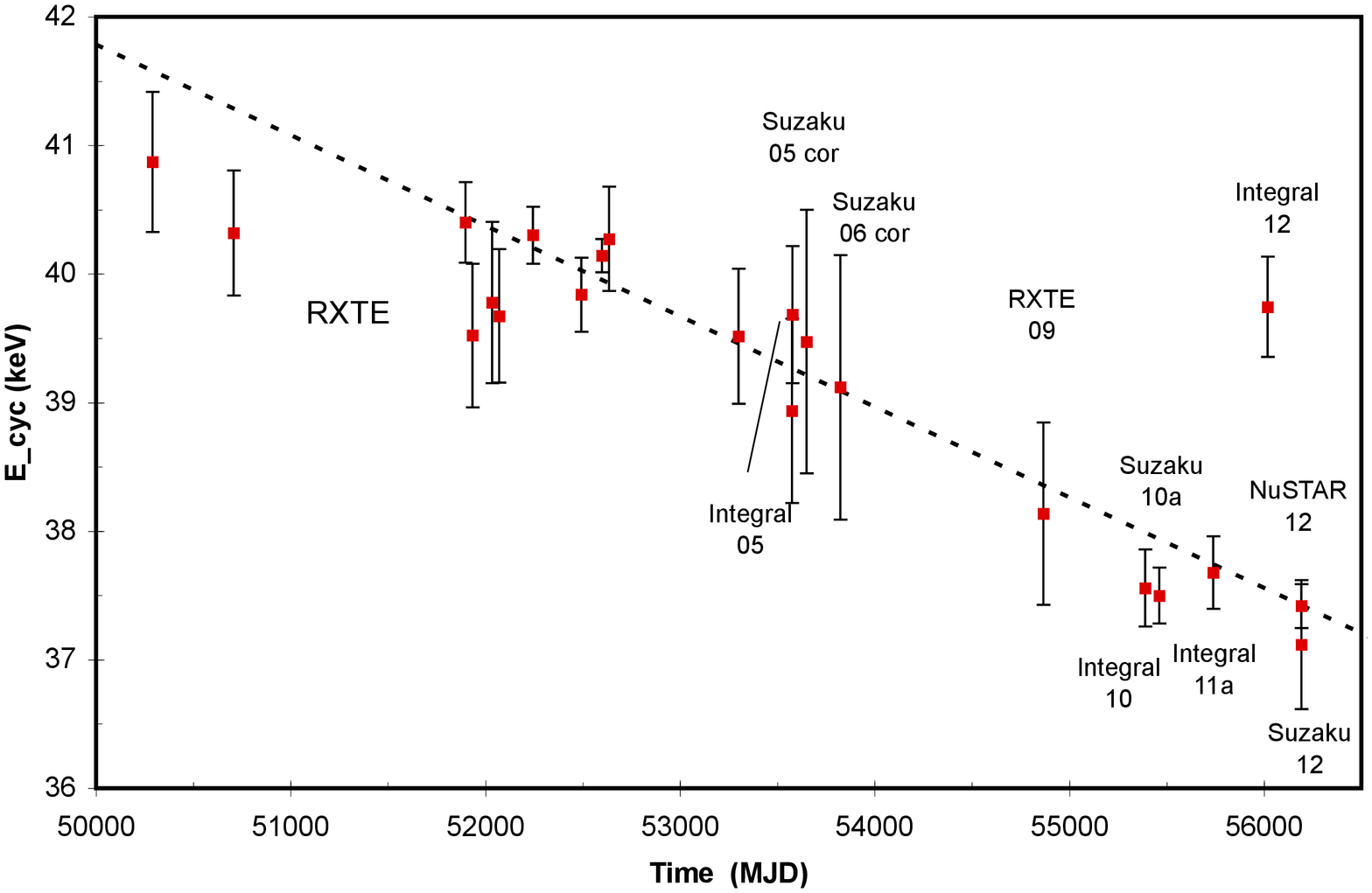}
\hfill
\vspace{-1cm}
\caption{Her~X-1 pulse phase averaged cyclotron line energies E$_{\rm cyc}$ normalized
  to a reference ASM count rate of 6.8\,cts/s using a flux dependence of 0.44\,keV/ASM-cts/s. 
  The data are now consistent with a linear decline of E$_{\rm cyc}$ with time 
  with a slope of $-7.22\times 10^{-4}$\,keV$d^{-1}$).
 }
 \label{fig:Fnorm3}
\end{figure*}

\begin{table*}
\caption[]{Details of fits with formula (1) to E$_{\rm cyc}$ values observed between
1996 and 2012 (excluding \textsl{INTEGRAL}12). The reference flux is F$_{0}$ = 6.8\,(ASM-cts/s) and the reference time is T$_{0}$ = MJD 53500.}
\vspace{-3mm}
\begin{center}
\begin{tabular}{lllllllll}
\hline\noalign{\smallskip}
Fit  & Param.                  & E$_{\rm 0}$         & a                          & b                           & $\chi^{2}$ & dof  & F-values        & chance         \\
No. & fitted                     & [keV]                    & [keV/ASM-cts/s]   & [$10^{-4}$ keV/d]  &                  &        &                       & probab.         \\
\hline\noalign{\smallskip}
1     & E$_{\rm 0}$          & $39.12\pm0.07$  & 0.00                    &  0.00                     & 430           &  20  &                        &                       \\  
2     & E$_{\rm 0}$, a      & $39.08\pm0.07$  & $0.74\pm0.09$   &  0.00                     & 357           &  19  & F(1-2) = 3.86  & 0.06               \\ 
3     & E$_{\rm 0}$, b      & $39.29\pm0.07$  & 0.00                    &  $-7.59\pm0.39$   &  45.0         &  19  & F(1-3) = 163   & 9 10$^{-11}$  \\   
4     & E$_{\rm 0}$, a, b  & $39.25\pm0.07$  & $0.44\pm0.09$   &  $-7.22\pm0.39$   &  20.4         &  18  & F(3-4) = 131   & 5 10$^{-5}$    \\   
\noalign{\smallskip}\hline
   \label{tab:3D_3}
\end{tabular}\end{center}
\vspace{-5mm}
The F-values are calculated according to the following formula:
${\rm F} = \Delta({\chi^{2}})/{\chi^{2}_{2}}$ $\times~$dof$(2)$. The last column gives the F-test probability that the
improvement in $\chi^{2}$ (by adding one additional free parameter) is just by chance. 
\end{table*}

In Table~\ref{tab:3D_2} we summarize results of fits to the 1996-2006 data set
with successive numbers of free parameters. Assuming no dependence at all
(a = b = 0.0), as well as dependence only on time (a = 0) leads to unacceptable 
fits with $\chi^{2}$/dof of 3.4 and 2.9, respectively. Allowing simultaneous dependence
on flux and on time leads to a good fit ($\chi^{2}$/dof = 0.6), representing a 
description of these data with a satisfactory separation of flux and time dependence.
It is again verified, that the flux dependence is the dominating effect, as clearly 
seen when the F-test is applied. \footnote{The respective improvements in $\chi^{2}$, 
when one additional free parameter is added, yields chance probabilities of 1.7\%
and $<10^{-10}$, respectively, for an improvement by chance, when as the third
free parameter the time dependence or the flux dependence is added.}


Allowing a simultaneous flux and time dependence, reduces the flux dependence slightly 
(as compared to neglecting the time dependence). If we now use the
flux dependence of 0.54 keV/(ASM-cts/s) for the normalization of the complete data set 
(1996-2012), we arrive at Fig.~\ref{fig:Fnorm2}. We note the following features: \\
1) The mean cyclotron line energies before and after 54500 ($\sim 2007$) are
$40.1\pm0.1$ and $37.6\pm0.1$, respectively. The difference is significant to $>17$
standard deviations (similar to the result of Section~\ref{sec:flux_norm1}, where the
flux normalization was done with the original slope of 0.66 keV/(ASM-cts/s)). \\
2) As noted before, there is a small downward trend between 1996 and 2006. 
Again, an added quadratic term for the flux normalization is not significant. \\
3) Because of the lack of measurements in 2007 and 2008 (at 35\,d phases $<0.20$), we cannot
distinguish between a fairly abrupt drop within this period and a smooth change
from the 2004-2006 period to the 2009-2012 period (possibly with a somewhat stronger 
decay than between 1996 and 2006). \\
4) The lower panel of Fig.~\ref{fig:Fnorm2} shows the 2-10\,keV X-ray flux 
(the maximum flux of each 35\,d cycle) from monitoring by \textsl{RXTE}/ASM and
\textsl{Swift}/BAT. The typical variation of the flux by a factor of $\sim2$ (as also evident 
from Fig.~\ref{fig:correlation}) is apparent. The mean flux, averaged over several 35\,d 
cycles, is constant.

\subsection{Normalizing E$_{\rm cyc}$ using fits to the 1996-2012 data with two variables}
\label{sec:flux-time_norm3}

We finally turn to fits with two variables (flux and time) to the data set of 1996-2012. 
For these fits we exclude the data point measured in April 2012 by \textsl{INTEGRAL}
since, as noted before, this value is not consistent with measurements by both
\textsl{Suzaku} and \textsl{NuSTAR} about six months later. 

As discussed in 
Section~\ref{sec:luminosity-dependence}, we have invested a strong effort to check the 
calibration of the \textsl{INTEGRAL}/IBIS detector for the time of observation and the 
subsequent data analysis procedure. We have not found any errors, so we keep this 
point in our data base (and show it in all plots), but exclude it from the fits to be discussed now
(but even if this data point is included, the general conclusion about the long-term decrease in the 
cyclotron line energy is not changed). As before, the bilinear function (formula (1)) is applied. 
The results of these fits with increasing numbers of free parameters are 
summarized in Table~\ref{tab:3D_3}, the fits are numbered 1 through 4. 
Using the flux dependence found in the final (simultaneous) fit (fit~4) for normalizing 
the observed E$_{\rm cyc}$ values to the reference flux of 6.8\,(ASM-cts/s), 
the remaining linear time dependence is shown in Fig.~\ref{fig:Fnorm3}.
The bilinear fit number 4 is acceptable with a $\chi^{2}$ = 20.4 for 18 dof.

From Fig.~\ref{fig:Fnorm3} we see that the \textsl{INTEGRAL}12 point is far ($\sim6$\,$\sigma$)
from the best fit (the dashed line). If we now repeat the bilinear fit including this point,
the linear time dependence is not changed significantly (from $(-7.22\pm0.39)$$\times10^{-4}$
to $(-6.91\pm0.39)$$\times10^{-4}$\,keV~d$^{-1}$), which means that the conclusions
about the decay of  E$_{\rm cyc}$ with time are not changed.

We note that a further improvement of the fit can be achieved by introducing a quadratic 
term in the time dependence (again for \textsl{INTEGRAL}12 excluded): 
the flux dependence is unchanged, the linear time term is now
$(-6.59\pm0.49)$$\times10^{-4}$\,keV~d$^{-1}$ and the quadratic time term is
$(-6.88\pm3.18)$$\times10^{-8}$\,keV~d$^{-2}$ (both with a reference time MJD~53500),
and a $\chi^{2}$ = 15.7 for 17 dof. However, the improvement is only marginally significant
with an F-value of 5.1, corresponding to a probability of 3.7\% for an improvement by chance.

Referring to Table~\ref{tab:3D_3} and Fig.~\ref{fig:Fnorm3} we note the following 
results: \\
1) Clearly, fit~1 is not acceptable ($\chi^{2}$ = 430 for 20 dof): E$_{\rm cyc}$ is not constant. \\
2) Introducing a linear flux dependence (neglecting any time dependence (fit~2)), improves the 
$\chi^{2}$ significantly and finds a flux dependence of $0.74\pm0.09$\,keV/(ASM-cts/s), which is 
(within uncertainties) consistent with the value found for the previous fit over the shorter time range 
(Table\ref{tab:3D_2}). However, a flux dependence alone is not sufficient. 
In addition, when a quadratic term for the flux normalization is introduced, assuming no time dependence, 
the fit is not acceptable ($\chi^{2}$ = 313, compare the corresponding values of Table \ref{tab:3D_3}). \\
3) Introducing a linear time dependence (neglecting any flux dependence (fit~3)), improves the
$\chi^{2}$ dramatically. The F-value corresponds to a formal probability of $<10^{-10}$ for the 
improvement to be just by chance. \\
4) Adding now the linear flux dependence (fit~4), $\chi^{2}$ is further reduced significantly, 
meaning that the flux dependence is definitely needed in addition to the time dependence (the F-value
corresponds to a chance probability of $\sim5\times10^{-5}$). The slope describing the flux
dependence is somewhat reduced as compared to fit~2, but is now very close to the corresponding
dependence found in fitting the $<2006$ data (see Table~\ref{tab:3D_2}). \\
5) We note that now the time dependence is the more dominant variation, while for the $<2006$ data 
set it was the flux dependence. This is consistent with a $\sim4$\,keV reduction in E$_{\rm cyc}$
over the covered time range (1996-2012), while the increase is only $\sim3$\,keV
over the flux range provided by nature (a factor of $\sim2$ in flux). A reduction in importance 
of the flux dependence, when adding the data after 2006, can already be expected from 
Fig.~\ref{fig:correlation}: the $>2006$ data are consistent with no flux dependence at all (even 
though they are formally consistent with the same slope as the $<2006$ data). \\
6) An even better fit can be achieved by introducing a quadratic term in the time dependence
(E$_{\rm cyc}$/dt$^{2}$ = $(-6.88\pm3.18)$$\times10^{-8}$\,keV~d$^{-2}$), albeit marginally
significant (3.7\% chance probability). The negative second derivative means that the decrease
in E$_{\rm cyc}$ accelerates with time. We note, that 
one might rather expect the opposite, e.g., a kind of exponential decay. \\
7) It is not understood why the measurement of \textsl{INTEGRAL} in 2012 is so different from
the nearby ($<$six months) observations  by \textsl{Suzaku} and \textsl{NuSTAR}. We find no errors
in our analysis. However, this one data point does not in any way change the general conclusion.

 \section{Summary of observational results}
\label{sec:summary}

Here we summarize our observational results:
\begin{enumerate}
\item The main result of this research is that we finally can establish a long-term
  decay of the pulse phase averaged cyclotron line energy E$_{\rm cyc}$ over time.
  The time range covered here is 16 years (1996 to 2012). The reduction is highly 
  significant and can be well described by a linear decay with a change by 
  $(7.2\pm0.4)$$\times10^{-4}$\,keV~d$^{-1}$ (0.26\,keV~yr$^{-1}$ or 4.2\,keV over 16\,yrs).
  The data are, however, also consistent with two other models: First (see Fig.~\ref{fig:Fnorm2}),
  a somewhat slower decay ($\sim$3$\times10^{-4}$\,keV~d$^{-1}$) until 2006 and then a 
  more sudden drop between 2006 and 2009, with a possible constant value ($\sim$37.7\,keV)
  thereafter. Alternatively, even an acceleration of the decay of E$_{\rm cyc}$ over time is possible, 
  since the fit with a quadratic term in the time dependence is formally the best one.
\item The flux dependence of E$_{\rm cyc}$ is confirmed with a value between 0.44 and 
  0.54\,keV/ASM-cts/s (with a typical uncertainty of 0.1), corresponding to a $\sim$5\%  increase
  for a factor of two increase in flux. This value is slightly lower than the value ($0.66\pm0.10$)
  from the original discovery \citep{Staubert_etal07} (at that time, a time dependence was neglected).
  It is not excluded that there is a variation with time of this flux dependence, but this is difficult to judge 
  since the fluxes observed after 2006 only cover a very small range.
  \item One observation, the one from \textsl{INTEGRAL} in April 2012, does not fit into the
  overall picture and is inconsistent with values measured by \textsl{Suzaku} and \textsl{NuSTAR}
  six months later. Since we find no errors in our analysis, we can only speculate that it is due to
  real physics -- that is a fluctuation on a timescale of a few months.
\item From the analysis of pulse profiles and pulse phase resolved spectroscopy 
  (see Section~\ref{sec:35d_phase}), both as a function of 35\,d phase, we find evidence of a variation 
  of E$_{\rm cyc}$ with this precessional phase: while there is very little (if any) variation
  until 35\,d phase 0.2, there may be a decrease at later phases. If it is indeed so, that the
  pulse phase dependence of E$_{\rm cyc}$ does not change with 35\,d phase \citep{Vasco_etal13},
  then this effect is inevitable. The fully covered \textsl{Main-On} of cycle 232 (Fig.~\ref{fig:35d})
  and the few dedicated observations at late 35\,d phases support this finding. 
  This new result adds another piece to the puzzle on the question about the physical nature 
  of the 35\,d modulation - precession of the accretion disk (plus?) precession of the neutron star? -
  and the mechanism of generating the varying pulse profiles and the varying spectra - continua and CRSF
  (see Section~\ref{sec:Discussion} and discussions in \citealt{Staubert_etal09, Staubert_etal13,Vasco_etal13}). 
\end{enumerate}

\vspace{-1mm}
\section{Discussion}
\label{sec:Discussion}

\subsection{Dependence of E$_{\rm cyc}$ on luminosity}
\label{sec:on_lum}

A negative correlation between the pulse phase averaged cyclotron line energy 
and the X-ray luminosity (a decrease in E$_{\rm cyc}$ with increasing $L_x$), 
had first been noted by \citet{Mihara_95} in observations of a few 
high luminosity transient sources (4U~0115+63, Cep~X-4, and V~0332+53) by \textsl{Ginga}. 
This negative correlation was associated with the high accretion rate during the X-ray 
outbursts, as due to a change in height of the shock (and emission) region above the 
surface of the neutron star with changing mass accretion rate, $\dot{M}$. 
In the model of \citet{Burnard_etal91}, the height of the polar accretion structure 
is tied to $\dot{M}$. From this model one expects that an increase in accretion rate leads 
to an increase in the height of the scattering region above the neutron star surface, and
therefore to a decrease in magnetic field strength and hence a decrease in E$_{\rm cyc}$. 
During the 2004/2005 outburst of V~0332+53 a clear anti-correlation of the line position 
with X-ray flux was observed \citep{Tsygankov_etal06}.
A similar behavior was observed in outbursts of 4U~0115+63 in March/April 1999
and Sep/Oct 2004: both \citet{Nakajima_etal06} and \citet{Tsygankov_etal07} had 
found a general anti-correlation between E$_{\rm cyc}$ and luminosity. However, 
\citet{SMueller_etal13}, analyzing data of a different outburst of this source in March/April
2008, observed by \textsl{RXTE} and \textsl{INTEGRAL}, have found that the negative 
correlation for the fundamental cyclotron line is likely an artifact due to correlations between 
continuum and line parameters when using the NPEX continuum model.

The first positive correlation was discovered by \citet{Staubert_etal07} in Her~X-1, 
and secured by a reanalysis of the same \textsl{RXTE} data 
by \citet{Vasco_etal11}, using the bolometric X-ray flux as reference. 
This analysis confirmed that the originally used 2-10\,keV flux is a good 
measure of the bolometric luminosity. While the above discussed analysis tests the correlated 
variability of E$_{\rm cyc}$ and $L_x$ on long timescales (35\,d and longer), the 
pulse-amplitude resolved analysis of \citet{Klochkov_etal11} does so on short timescales 
(down to the pulse period of 1.24\,s). Selecting pulses with amplitudes in certain ranges and 
producing mean spectra for each pulse amplitude range, showed that the cyclotron line energy 
scales positively with the mean pulse amplitude. In addition, it was found that the photon index 
$\Gamma$ of the underlying power law continuum scales negatively with the pulse amplitude
(the absolute value of $\Gamma$ gets smaller, that is the spectrum flattens).
The same behavior was seen in data of the transient A~0535+26. A recent pulse phase resolved 
analysis of A~0535+26 observations by \textsl{RXTE} and \textsl{INTEGRAL} showed that data 
of one of the two peaks (of the double peak pulse profile) displays the same trend while data of
the other peak do not \citep{DMueller_etal12,DMueller_etal13}.
Applying the same pulse-amplitude resolved technique to data of V~0332+53 and 
4U~0115+63, \citet{Klochkov_etal11} found the same behavior as originally detected in
data sets that were selected on much longer timescales: E$_{\rm cyc}$ decreases and 
$\Gamma$ increases with increasing $L_x$.
Finally, we mention that a positive correlation of E$_{\rm cyc}$ with $L_x$ was also found
in two more X-ray binary pulsars: in GX~304-1 \citep{Yamamoto_etal11,Klochkov_etal12}
and in \textsl{NuSTAR} observations of Vela~X-1 \citep{Fuerst_etal14}. We note, that the
still small group of four objects with a positive E$_{\rm cyc}$/$L_x$ correlation now outnumbers
the group of secure sources with the originally discovered opposite behavior.

Our current understanding of the physics behind these correlations assumes that we can
distinguish between \textsl{two accretion regimes} in the accretion column above the polar cap
of the neutron star: \textsl{super- and sub-Eddington accretion}. The former is responsible for
the first detected negative correlation in high luminosity outbursts of transient X-ray sources 
(the reference source being V~0332+53): in this case the deceleration of the accreted 
material is provided by radiation pressure, such that with increasing accretion rate 
$\dot{M}$, the shock and the scattering region move to larger height above the surface of 
the neutron star and consequently to weaker B-field \citep{Burnard_etal91}. 
Sub-Eddington accretion, on the other hand, leads to the opposite behavior. In this regime 
the deceleration of accreted material is predominantly through Coulomb interactions
and an increase in $\dot{M}$ leads to an increase in electron density (due to an increase
of the combined hydrostatic and dynamical pressure) resulting in a \textsl{squeezing}
of the decelerating plasma layer to smaller height and stronger B-field \citep{Staubert_etal07}.
More detailed physical considerations have recently been presented by \citet{Becker_etal12}.
The persistent sources Her~X-1 and Vela~X-1are clearly sub-Eddington sources.

Despite the above discussed doubts about the reality of the negative correlation of the energy E$_{\rm cyc}$ 
of the fundamental CRSF with $L_x$ in 4U~0115+63 \citep{SMueller_etal13}, and keeping in mind that
\citet{Klochkov_etal11} had confirmed the correlation using the pulse-amplitude resolved technique
(not using the NPEX function), we would like to note here that we are intrigued by the following plots about
4U~0115+63: Fig.~8 of  \citet{Nakajima_etal06} and Figs.~11 and 12 of \citet{Tsygankov_etal07}. 
In going to the lowest luminosities, there is an indication for a leveling-off or even a reduction in 
E$_{\rm cyc}$. Do we possibly see here the transition between the two accretion regimes?

\subsection{The long-term decay of E$_{\rm cyc}$}
\label{sec:on_time}


With regard to the physical interpretation of the now observed long-term decrease in the 
cyclotron line energy, we speculate that it could be connected to either unknown effects in
the neutron star and its magnetic field, to a geometric displacement of the cyclotron resonant 
scattering region in the dipole field or to a true physical change in the magnetic field configuration
at the polar cap, which evolves due to continued accretion. Apparently, the magnetic
field strength at the place of the resonant scattering of photons trying to escape from the
accretion mound surface must have changed with time. Putting internal neutron star physics aside,
we suggest that it reflects a local phenomenon in the accretion mound: either a geometric 
displacement of the emission region or a change in the local field configuration, rather than a change 
in the strength of the underlying global dipole field (here a minimum timescale of a million years 
is estimated from population studies of rotation-powered pulsars; \citealt{Bhattacharya_etal92,GeppertUrpin_94}).
Our observed timescale, a few tens of years, is extremely short.

The whole issue of accretion onto highly magnetized neutron stars in binary X-ray sources is 
very complex. Ideas or models with potential relevance to our observations attempting
to understand the magnetic field configuration in accreting neutron stars and its evolution over 
extended periods of continued accretion, can be found in e.g.:
\citet{Hameury_etal83,KonarBhatt_97,BrownBildsten_98,ChengZhang_98,Litwin_etal01,
Cumming_etal01,MelatosPhinney_01,ChoudhuriKonar_02,PayneMelatos_04,PayneMelatos_07,
Wette_etal10,MukhBhatt_12,Mukherjee_etal13a,Mukherjee_etal13b}.
However, as far as we can see, none of them gives the complete picture. Most calculations 
deal with static solutions that are found under special boundary conditions. 

Since the main purpose of this contribution is to report on the discovery of a new observational
phenomenon, we refrain from going into any details regarding interpretations of
existing models.
Instead, we only mention a few areas which we think could have 
some connection with the observed facts and which may be worthwhile to be explored.
Our hope is that the new observational results presented here may boost the motivation for further 
theoretical studies.

We start by asking whether the observed decrease in E$_{\rm cyc}$ with time could be a
simple movement of the resonant scattering region to a larger distance from the neutron star
surface, where the field strength is lower. This would be similar to the decrease in E$_{\rm cyc}$
during outbursts in high luminosity transients, except that (being in the sub-Eddington regime
of accretion) we would not think of the shock region to rise, but rather the total height of the
accretion mound may slowly increase with time, such that also the resonant scattering region is
displaced to a higher position. For a dipole field with an r$^{-3}$ dependence of field strength,
the observed $\sim$5\,keV reduction in E$_{\rm cyc}$ from 1992 to 2012 (0.25\,keV per year)
would correspond to a change in height of $\sim$400\,m (starting from the surface itself).
The question here is, whether continued accretion really leads to a growth of the
accretion mound with time - both in terms of geometrical height and of total mass.

With similar uncertainty, it can be asked whether the accreted material
could drag the central field lines radially out, possibly enlarging the total hotspot area and thereby
diluting the effective field strength in the region where the resonant scattering takes place 
\citep{ChengZhang_98,ZhangKojima_06}. Or, whether \textsl{screening} or \textsl{burial}
of the magnetic field at the polar caps is possible 
\citep{BrownBildsten_98,PayneMelatos_04,Litwin_etal01,PayneMelatos_07}. It needs to be
investigated, how much mass could eventually be stored in the magnetically confined mountains,
whether matter is continuously leaking out to larger areas of the neutron star surface
(due to plasma pressure exceeding the magnetic pressure)
and on what timescales an observational effect can be expected.

Finally, the question of Ohmic dissipation and diffusion of the magnetic field
may play a role and physical processes either in the accretion mound  or in or below
the surface of the neutron star (like hydrodynamic flows) could \textsl{bury} or reduce the surface field 
\citep{ChoudhuriKonar_02,Patruno_12}. One would need to investigate whether
physical parameters like the characteristic length scale and the relevant conductivity
$\sigma$ for either the crust or the plasma in the accretion mound could be of the
right order of magnitude to allow the magnetic diffusion timescale 
$\tau$ = 4$\pi$R$^{2}$$\sigma$/c$^{2}$ \citep{Cumming_etal01, Ho_11} to be
compatible with the timescale of a few tens of years - as observed for the decrease
in the local polar field strength in Her X-1.
If magnetic diffusion is indeed relevant, we note that the necessary small length scales 
and relatively low conductivities would argue for local physics in the hot plasma of the 
accretion mound, the structure of which is most likely complex because of contributions
from higher-order multipoles.

We finally speculate on a possible cyclic behavior of E$_{\rm cyc}$ on timescales
of a few tens to hundreds of years. Could it be that the fast rise of the observed E$_{\rm cyc}$
values after 1991 (see Fig.~\ref{fig:history}) represents a special event in which the magnetic
field in the accretion mound has rearranged itself as a result of a sudden radial
outflow of material? In models by \citet{BrownBildsten_98,Litwin_etal01,PayneMelatos_04,
PayneMelatos_07,MukhBhatt_12,Mukherjee_etal13a} the field configuration is shown to change
considerably with increased material, leading to a \textsl{ballooning} of the field configuration with
diluted field in the symmetry center and increased density of field lines at the circumference
of the base of the mound. The estimates of how much mass could be confined by
the field vary substantially between the different models. It remains unclear, how important
continuous leaking through the outer magnetic boundary may be and what the timescales
for semi-catastrophic events might be, in which the field would release 
(on a short timescale) a substantial fraction
of stored material to larger areas of the neutron star surface. For Her~X-1 this scenario could 
mean that we are now in a phase of continuous build-up of the accretion mound with
the mass (and the height?) of the mound growing and the observed cyclotron line energy 
continuously decreasing until another event like the one around 1991 happens again. The
mean E$_{\rm cyc}$ value measured before 1991 of $\sim$35\,keV may represent a bottom
value. So, when the current decay continues steadily, one may expect another event of
a rather fast increase in E$_{\rm cyc}$. 

In conclusion, we like to urge both observers and model builders to continue to accumulate
more observational data as well as more understanding of the physics responsible for
the various observed properties of Her~X-1 and other objects of similar nature. 
For model builders a challenge would be to work towards
dynamical computations that might eventually lead to self-consistent solutions of the structure and 
evolution of magnetized accretion mounds of accreting neutron stars with only a few input parameters. 

\begin{acknowledgements}
This paper is to a large part based on observational data taken by the NASA satellite 
\textsl{Rossi X-Ray Timing Explorer} (RXTE). We like to acknowledge the dedication 
of all people who have contributed to the great success of this mission. In the same
way, we thank the teams of ESA's \textsl{INTErnational Gamma-Ray Astrophysics Laboratory} 
(INTEGRAL), JAXA's \textsl{Suzaku} and NASA's \textsl{Nuclear Spectroscopic Telescope Array} 
(NuSTAR).
This work was supported by DFG through grants Sta~173/31-1,2 and
436~RUS~113/717 and RFBR grants RFFI-NNIO-03-02-04003 and 06-02-16025.
The work of K.P. and N.Sh. was also partially supported by RBFR grants 12-02-00186 and 14-02-00657.
D.K. is indebted to the Carl Zeiss Stiftung for support.
We thankfully acknowledge very useful discussions about the possible physical
meaning of the observed effects with D. Bhattacharya, K. Kokkotas, K. Glampedakis
and J. Tr\"umper. Finally we thank the anonymous referee for important questions and
suggestions.
\end{acknowledgements}

\bibliographystyle{aa}
\bibliography{refs_herx1}

\end{document}